\documentclass[aps,twocolumn,showpacs,preprintnumbers,nofootinbib,prd,superscriptaddress,10pt]{revtex4-2}

\makeatletter
\def\l@subsubsection#1#2{}
\def\l@subsubsubsection#1#2{}
\makeatother

\usepackage{comment}
\usepackage{graphicx,amssymb,amsmath,amsthm,amsfonts,epsfig,epsf,fixmath}
\usepackage[usenames]{color}
\usepackage{xcolor}
\usepackage{epstopdf}

\usepackage{aas_macros}
\usepackage{bm}
\usepackage{dcolumn}
\usepackage{lipsum}
\usepackage{latexsym}
\usepackage{rotating}
\usepackage{longtable}

\usepackage{enumerate}
\usepackage{tensor,multirow}
\usepackage{url}
\usepackage[linktocpage]{hyperref}

\begin{document}

\title{Superradiant instabilities by accretion disks in scalar-tensor theories}

\author{Giuseppe Lingetti}
\affiliation{Dipartimento di Fisica, ``Sapienza'' Universit\`a di Roma \& Sezione INFN Roma1, Piazzale Aldo Moro 
5, 00185, Roma, Italy}
\author{Enrico Cannizzaro}
\affiliation{Dipartimento di Fisica, ``Sapienza'' Universit\`a di Roma \& Sezione INFN Roma1, Piazzale Aldo Moro 
5, 00185, Roma, Italy}
\author{Paolo Pani}
\affiliation{Dipartimento di Fisica, ``Sapienza'' Universit\`a di Roma \& Sezione INFN Roma1, Piazzale Aldo Moro
5, 00185, Roma, Italy}

\begin{abstract} 
We study the superradiant instability in scalar-tensor theories of gravitation, where matter outside a black hole provides an effective mass to the scalar degree of freedom of the gravitational sector. We discuss this effect for arbitrarily spinning black holes and for realistic models of truncated thin and thick accretion disks (where the perturbation equations are nonseparable), paying particular attention to the role of hot coronal flows in the vicinity of the black hole. The system qualitatively resembles the phenomenology of plasma-driven superradiant instabilities in General Relativity. Nevertheless, we show that the obstacles hampering the efficiency of plasma-driven superradiant instabilities in General Relativity can be circumvented in scalar-tensor theories.  We find a wide range of parameter space where superradiant instabilities can be triggered in realistic scenarios, and discuss the constraints on scalar-tensor theories imposed by this effect.
In particular, we argue that the existence of highly spinning accreting black holes is in tension with some scalar-tensor alternatives to the dark energy, e.g. symmetron models with screening.
\end{abstract}

\maketitle

\section{Introduction}
\subsection{Motivation}
Scalar-tensor theories are among the most interesting and well-studied extensions of General Relativity~(GR). In this class of theories, the gravitational sector includes one or more scalar fields which are nonminimally coupled to the standard metric. A quite general action of scalar-tensor theories with one scalar field reads~\cite{Fujii:2003pa}:
\begin{multline}
\label{eq:jordanaction}
    S= \frac{1}{16\pi G}\int d^4x \sqrt{-g}[\mathcal{F}(\phi)R-Z(\phi)g^{\mu\nu}\partial_\mu \phi \partial_\nu \phi-\\-U(\phi)]+S_m(\psi_m, g_{\mu\nu})\,,
\end{multline}
where $R$ is the Ricci scalar, $g_{\mu\nu}$ is the metric, $\phi$ is a scalar field, and the last term denotes the action of matter fields minimally coupled to the metric. Depending on the expressions of the functions $\mathcal{F}$, $Z$, and $U$, it is possible to recover different theories. For example, for $\mathcal{F}=\phi$, $Z\propto\phi^{-1}$, and $U=0$, Eq.~\eqref{eq:jordanaction} represents Brans-Dicke theory. Actions with scalar fields nonminimally coupled to gravity also arise from string theory, Kaluza-Klein-like theories, and braneworld scenarios. These theories have been intensively investigated in cosmology~\cite{Faraoni:2004pi, Clifton:2011jh}. Likewise, astrophysical implications of scalar-tensor theories for compact objects have been explored in detail~\cite{Berti:2015itd}.

A crucial requirement for these theories is that their weak-field limit, i.e. length scales between the micrometer and the astronomical unit, must be consistent with GR, which in this regime has been tested with extreme precision~\cite{Will:2014kxa,Berti:2015itd}. Typically, scalar-tensor theories with interesting cosmological phenomenology must feature some screening mechanism, hiding the scalar field on local scales~\cite{Hinterbichler:2010es, Khoury:2003aq}. 
It is thus relevant to study the phenomenology of these theories in the strong gravity regime, where deviations from GR might be more dramatic.
In this work, we perform a detailed analysis of matter-triggered superradiant instabilities for spinning black holes~(BHs) in scalar-tensor theories (see~\cite{Brito:2015oca} for an overview on BH superradiance). This effect was unveiled in~\cite{Cardoso:2013opa, Cardoso:2013fwa}, where it was shown that the presence of matter outside BHs can trigger either spontaneous scalarization or a superradiant instability in the system, due to the scalar field acquiring an effective mass squared proportional to the trace of the stress-energy tensor of the surrounding matter. The scope of this work is to investigate whether this superradiant instabilities can arise if one considers realistic models of accreting BHs. A similar analysis was recently performed in~\cite{Dima:2020rzg} in the context of plasma-driven~\cite{Pani:2013hpa,Conlon:2017hhi} superradiant instabilities of photons in GR for BHs accreting a tenuous plasma, using a spin-0 toy model  (see also~\cite{Wang:2022hra} for an extension to the Proca case, and~\cite{Cannizzaro:2020uap, Cannizzaro:2021zbp} for a recent analysis of photon-plasma interactions in curved spacetime). It was shown in~\cite{Dima:2020rzg} that the complex geometry of accretion disks and the high values of plasma density near the BH can significantly quench the instability.

Nevertheless, we show that this problem can be circumvented in scalar-tensor theories for realistic accretion-disk configurations, because the effective mass depends also on the scalar-tensor coupling. 
For a cold, collisionless plasma the effective photon mass corresponds to the plasma frequency~\cite{Pani:2013hpa,Conlon:2017hhi,Cannizzaro:2020uap, Cannizzaro:2021zbp}:
\begin{equation}
    \omega_p=\sqrt{\frac{4\pi e^2 n_e }{m_e}}\approx 10^{-12}\sqrt{\frac{n_e}{10^{-3}{\rm cm}^{-3}}}\,{\rm eV}\,,
\end{equation}
where $n_e$ is the number density of the free electrons (with mass $m_e$ and charge $e$) in the plasma. 
BH superradiant instabilities are most effective when the gravitational coupling $\omega_p M\sim {\cal O}(0.1)$, where $M$ is the BH mass, and highly suppressed if $\omega_p M\gg1$. For astrophysically relevant BHs with $M> M_\odot$, this condition on the coupling implies $\omega_p\lesssim 10^{-11}\,{\rm eV}$. Thus, the effective mass lies in a range able to trigger superradiant instabilities if $n_e\sim 10^{-3}-10^{-2}\,{\rm cm}^{-3}$, i.e. for plasma densities typical of the interstellar medium~\cite{Conlon:2017hhi}.
The plasma density near an accreting BH is several orders of magnitude bigger~\cite{Dima:2020rzg}. In this case, the effective mass is too large to induce an instability on a sufficiently short time scale. 

However, as we shall later discuss, in scalar-tensor theories the effective mass squared is~\cite{Cardoso:2013opa, Cardoso:2013fwa}
\begin{equation}
    \mu_{\rm eff}^2=-2\alpha T\sim 2{\alpha} \rho \,,
\end{equation}
where $T$ is the trace of the stress-energy tensor, $\rho=m_N n_e$ is the matter-energy density of the gas (with nucleon mass $m_N$), ${\alpha}$ is a free parameter related to the nonminimal coupling of the scalar field, and the last step above is valid for a nonrelativistic disk (see details below). Thus, in the scalar-tensor case the effective mass depends on $n_e^{1/2}$ as in the standard photon-plasma case but, crucially, also on a free effective coupling ${\alpha}$. As we shall discuss, depending on the value of ${\alpha}$, the effective mass can be in the optimal range to trigger a superradiant instabilities for realistic plasma configurations around BHs.

Another effect that can drastically quench plasma-driven BH superradiant instabilities are nonlinearities~\cite{Cardoso:2020nst}. While transverse waves with frequency $\omega<\omega_p$ do not propagate in a cold plasma within linear theory, nonlinear effects make the plasma transparent if the electric field is higher than $E_{\rm crit}=\frac{m_e}{e}\sqrt{\omega_p^2-\omega^2}$~\cite{1970PhFl...13..472K,1971PhRvL..27.1342M}. This effectively corresponds to the fact that the plasma frequency is decreased by a Lorentz boost factor arising from the backreaction of the plasma four-velocity. During the superradiant growth of the electric field the Lorentz factor can be significantly large, severely limiting the angular momentum and energy extraction through plasma-driven superradiant instabilities within GR~\cite{Cardoso:2020nst}. 
As we shall later discuss, the situation is radically different in the case of scalar-tensor theory. Also in this case the backreaction induces a change in the plasma four-velocity but, because the effective mass depends only on the trace of the stress-energy tensor, it is not suppressed by a Lorentz factor.

Throughout this paper, we use $G=c=1$ units and the $(-,+,+,+)$ signature.

\section{Setup}
\label{sec:setup}

\subsection{General equations and framework}
\label{sub:general equation}

The action in Eq.~\eqref{eq:jordanaction} is in the so-called Jordan frame, where the scalar field is nonminimally coupled to the metric.
By performing a conformal transformation of the metric and a field redefinition for the scalar field,
\begin{align}
    &g_{\mu\nu}^E=\mathcal{F}(\phi)g_{\mu\nu}\,, \\&
    \Phi(\phi)=\frac{1}{4 \pi}\int d\phi \Bigg[\frac{3}{4}\frac{\mathcal{F}'(\phi)^2}{\mathcal{F}(\phi)^2}+\frac{1}{2}\frac{Z(\phi)}{\mathcal{F}(\phi)}\Bigg]^{1/2}\nonumber \,,\\&
    A(\Phi)=\mathcal{F}^{-1/2}(\phi)\nonumber \,,\\&
    V(\Phi)=\frac{U(\phi)}{\mathcal{F}^2(\phi)}\,, \nonumber
\end{align}
it is possible to describe the system in the so-called Einstein frame, where the action takes the form:
\begin{multline}
    S=\int d^4x \sqrt{-g^E}\Bigg(\frac{R^E}{16\pi}-\frac{1}{2}g^E_{\mu\nu}\partial^\mu \Phi \partial^\nu \Phi-\frac{V(\Phi)}{16 \pi}\Bigg)\\+ S(\psi_m, \mathcal{A}(\Phi)^2g^E_{\mu\nu})\,.
\end{multline}
In the Einstein frame, the scalar field is minimally coupled to the gravity sector, but matter fields are coupled to the effective metric $A(\Phi)^2g^E_{\mu\nu}$, so that the weak equivalence principle is preserved while its strong version is violated.
In this frame, we assume a generic analytic behavior for the potentials around a GR solution with a constant value $\Phi^{(0)}$ of the scalar field\footnote{We consider the field equations in the Einstein frame but laboratory clocks and rods refer to the Jordan-field metric $g_{\mu\nu}= A^2 g_{\mu\nu}^2$. Physical asymptotic quantities related to the metric (e.g., masses and angular momementa) are obtained from their Einstein-frame counterpart by rescaling the latter with suitable powers of $A(\Phi^{(0)})$. In practice, recovering GR in the weak-field regime requires $A(\Phi^{(0)})\approx1$ so the distinction between Einstein- and Jordan-frame asymptotic quantities is negligible for our purposes.
},
\begin{align}
    V&=\sum_{n=0} V_n (\Phi-\Phi^{(0)})^n\, ,\\
    A&=\sum_{n=0} A_n (\Phi-\Phi^{(0)})^n\,.
\end{align}
Then, by expanding the field equations for $\varphi\equiv\Phi-\Phi^{(0)}\ll1$, it is possible to rearrange the field equation for the scalar field in a GR background as (see~\cite{Cardoso:2013opa, Cardoso:2013fwa} for details)
\begin{equation}
\label{eq:finalKGeq}
    [\Box^E-\mu_{\rm eff}^2(r, \theta)]\varphi=0\,,
\end{equation}
with an \emph{effective} mass squared term
\begin{equation}
    \mu_{\rm eff}^2(r, \theta)=\frac{V_2}{8\pi}-2{\alpha} T^E(r, \theta)\, ,
\end{equation}
where ${\alpha}=A_2/A_0$ 
Following~\cite{Cardoso:2013fwa} we focus on asymptotically-flat spacetimes (which requires $V_0=V_1=0$) and on theories admitting GR vacuum solutions (which requires $A_1=0$).
For the rest of this analysis we will also assume $V_2=0$. This term is related to a standard bare mass, and neglecting it corresponds to assuming a massless field. 

We are therefore left with a Klein-Gordon equation with an effective mass squared proportional to the trace of the stress-energy tensor of the surrounding matter. Since the matter backreaction on the metric is typically negligible, and 
owing to BH no-hair theorems in this class of theories~\cite{Sotiriou:2011dz}, the background is described by the Kerr solution.
The sign of the parameter ${\alpha}$ has a crucial impact on the phenomenology of the system~\cite{Cardoso:2013opa, Cardoso:2013fwa}. If ${\alpha}<0$ the effective mass squared in Eq.~\eqref{eq:finalKGeq} is negative, and leads to a possible tachyonic instability and to a scalarization of the BH. If instead ${\alpha}>0$, the effective mass squared is positive and the system can undergo a superradiant instability. In this work we are interested in the latter case.

Indeed, it is well known that spinning compact objects are unstable against massive bosonic degrees of freedom (see~\cite{Brito:2015oca} for an overview). For a bosonic field with mass term $\mu\lesssim 0.1/M$, the spectrum is approximately hydrogenic and modes are unstable when their frequency $\omega_R\approx \mu$ satisfies the superradiance condition $0<\omega_R<m \Omega_H$, where $\Omega_H$ is the BH angular velocity and $m$ is the azimuthal number of the mode. As a result of this instability, a macroscopic bosonic condensate forms around the BH, extracting energy and angular momentum from the latter.
The same physical effect occurs if the bosonic field possesses an effective mass, although in such case the instability depends also on the geometry of the effective-mass term, as we shall discuss.

\subsection{Effective mass}
\subsubsection{Stress-energy tensor of accretion disks}

As previously discussed the effective mass-squared term depends on the trace of the stress-energy tensor of the matter fields surrounding the BH. In this section we characterize this term for realistic accretion disk profiles.

We consider different types of effective mass. In general, the stress energy-tensor of an accretion disk can be fully described by four different components~\cite{Abramowicz2013}:
\begin{equation}
    T^\mu_\nu=(T^\mu_\nu)_{\rm FLU}+(T^\mu_\nu)_{\rm VIS}+(T^\mu_\nu)_{\rm MAX}+(T^\mu_\nu)_{\rm RAD}\, ,
\end{equation}
which are, respectively, the fluid component, the viscosity component, the electromagnetic component, and the radiation one. Most models of accretion disks assume a particular form of the stress energy-tensor. For example, thick accretion disk models rely on a perfect fluid approximation, which states that $(T^\mu_\nu)_{\rm VIS}=(T^\mu_\nu)_{\rm MAX}=(T^\mu_\nu)_{\rm RAD}=0$. Throughout this work, we will consider this assumption, in which the stress energy-tensor reads
\begin{equation}
     (T^\mu_\nu)_{\rm FLU}=(\rho u^\mu)(W u_\nu)+\delta^\mu_\nu P\, ,
\end{equation}
where $\rho$, $W$, $P$ are respectively the mass-energy density, enthalpy, and pressure. By neglecting the internal energy density of the fluid, the stress-energy tensor trace reads $T=-\rho+3P$. Note that while the perfect fluid approximation holds for thick disks, in our case we can use the same approximation also for thin disks. Thin disks have a nonvanishing stress part, which for example in the Shakura-Sunyev model can be described using a nearly-linear viscosity approximation~\cite{1973A&A....24..337S}. However, the stress part can be written as $(T^\mu_\nu)_{\rm VIS}\propto\sigma^{\mu}_{\nu}$, where $\sigma^{\mu\nu}$ is the shear tensor of the four-velocity of the fluid. Since the shear tensor is by definition \emph{traceless}, the effective mass is independent of the viscosity. 

In what follows we will also neglect the effect of pressure, as it is subdominant. Indeed, if one for example assumes the equation of state of an ideal gas, then $P=c_s^2 \rho$, where $c_s$ is the speed of sound of the fluid. Since for accretion disks $c_s$ is at least two orders of magnitude smaller than the speed of light, we are in the nonrelativistic regime, $P\ll\rho$, and we can safely neglect pressure corrections to the effective mass.
Thus, the trace of the stress-energy tensor in our models is simply $T\approx-\rho$.

\subsubsection{Accretion disks features: truncation, typical densities, and coronae}
In the following, we will be interested in accretion environments that exhibit a sharp cut-off sufficiently far away from the BH horizon.
In these models the disk creates a cavity that can potentially trap scalar modes leading to an instability. A system that satisfies this requirement is the truncated disk accretion model. Truncated disk models are commonly used in BH accretion physics and, depending on the accretion rate, the location of the truncation can be close to the Innermost Stable Circular Orbit (ISCO) (high/soft state) or very far from it, even at $200-400M$ or more (low/hard state). 
Whenever this happens, in the region within the truncation radius and down to the vicinity of the BH, only a hot coronal flow can exist (see e.g.~\cite{truncation2007, truncation2014, truncation2014b, truncation1997, truncation2013}).
The Comptonization of hot electrons in the coronal medium is believed to explain the hard, X-ray tail that follows the black-body like emission spectrum of the disk. For this reason, the truncated disk + corona model succeeds in explaining features in the emission spectrum~\cite{truncation2014}. 

Another ideal configuration producing sufficiently wide cavities in the density profile near the BH are counter-rotating disks that extend all the way to the ISCO. In this case, the ISCO is sufficiently far away from the horizon ($6\leq r_{\rm ISCO}/M\leq9$ depending on the BH spin) so that the cavity is able to trap modes. Finally, another interesting possibility are magnetically-arrested disks, where a strong poloidal magnetic field disrupts the disk at a relatively large radius, creating a cavity. Also this model supports the presence, inside the cavity, of a hot, low-density coronal flow \cite{Narayan:2003by}. In general, these flows are always very tenuous and quasispherical, and their density is lower than the disk's one by some orders of magnitude (see~\cite{Bisnovatyi-Kogan:1976fbc} for an estimate or, e.g.,~\cite{DeVilliers:2003gr, DeVilliers:2004zz} for GR magneto-hydrodynamics simulations). In what follows, we shall therefore describe truncated thin and thick disks by taking into account an additional coronal structure. 

\subsubsection{Plasma profiles}

We consider different models of density profiles, discussed below.
In all models, since the time scales of interest are much shorter than the typical BH accretion time scales~\cite{Cannizzaro:2021zbp}, we shall neglect the time dependence of the matter fields. Moreover, we shall restrict to axisymmetric configurations in which $\rho=\rho(r,\theta)$ (that of course reduce to spherical configurations for purely radial profiles). 

Model~I describes a thick disk+corona system where the corona is described by a constant asymptotic term. The full profile reads
\begin{equation}
     \mu_{\rm eff, I}^2(r,\theta)={\alpha}\left[\rho_H\Theta(r-r_0)\left(1-\frac{r_0}{r}\right)\left(\frac{r_0}{r}\right)^{\frac{3}{2}}+\rho_C\right]\,,
\end{equation}
where $\Theta(x)$ is the Heaviside step function. 
When the scalar coupling ${\alpha}=1$, this model coincides with the one studied in~\cite{Dima:2020rzg} with a suitable choice of the parameters $\rho_H, \rho_C$, and $r_0$.
In order to investigate the role of the mass at spatial infinity, in Model~II we truncate the corona at $r_0$:
\begin{multline}
     \mu_{\rm eff, II}^2(r,\theta)={\alpha}\left[\rho_H\Theta(r-r_0)\left(1-\frac{r_0}{r}\right)\left(\frac{r_0}{r}\right)^{\frac{3}{2}}\right.\\+\rho_C\Theta(r_0-r)\Big]\,.\label{model2}
\end{multline}
In Model~III we investigate the effects of the sharp cut-off produced by the Heaviside function in Models~I~and~II by replacing it with a sigmoid-like function:
\begin{equation}
     \mu_{\rm eff, III}^2(r,\theta)=\frac{{\alpha}\rho_H}{1+e^{-2(r-r_0)}}\left[1-\frac{r_0}{r\left(1+\frac{\beta}{r^4}\right)}\right] \left(\frac{r_0}{r}\right)^{\frac{3}{2}}\, .
\end{equation}

\begin{figure}[t]
\centering
\includegraphics[width=0.49\textwidth]{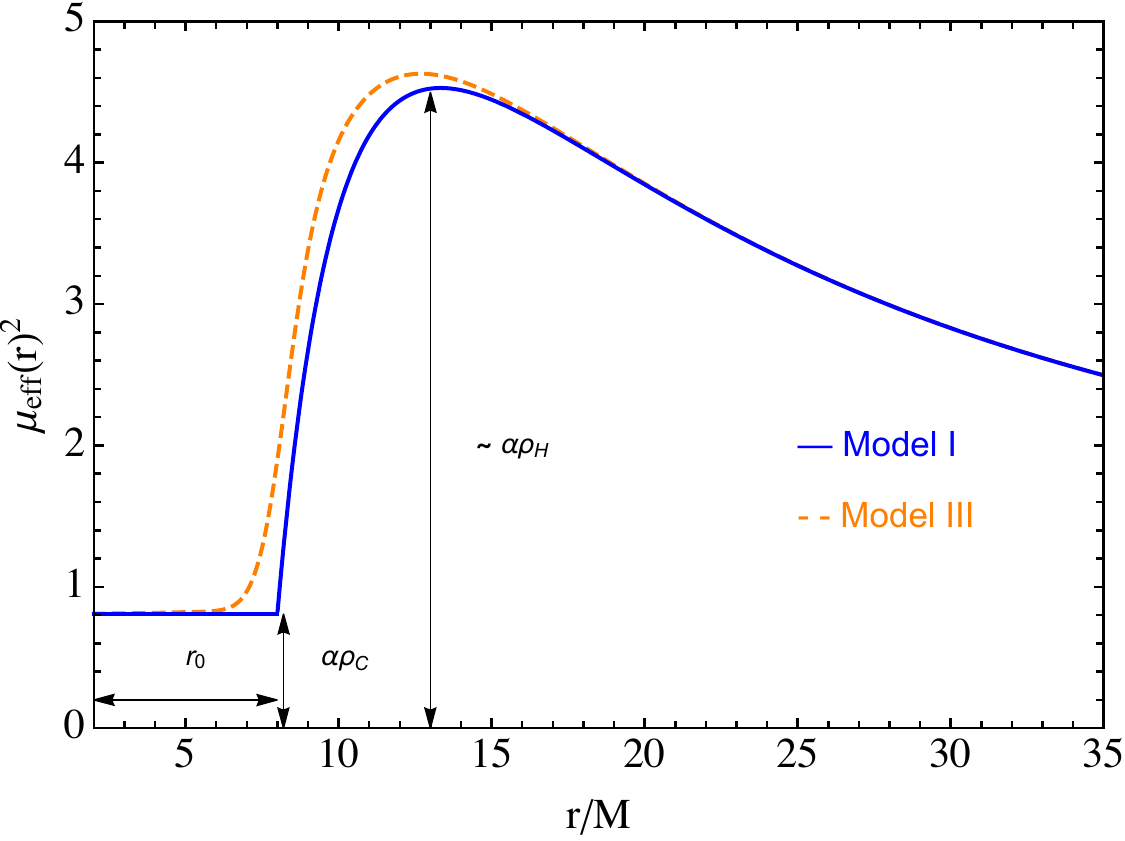}
\caption{Radial profile of the effective mass in Model~I with $\alpha\rho_C M^2=0.9$, $\alpha\rho_H M^2=20$ and $r_0=8M$ (solid blue) and Model~III with $\beta=500$ (dashed orange). The profiles are similar, but in Model~III the sharp cutoff is smoothed out. For convenience, we have chosen unrealistic values to better highlight the three fundamental parameters ($r_0,\alpha\rho_C, \alpha \rho_H$) that govern the salient features of the geometry.}
\label{fig:confrontation}
\end{figure}

Figure~\ref{fig:confrontation} shows that, with a suitable choice of $\beta$, Model~III is very similar to Model~I, except that the effective mass does not display a sharp cutoff. 

Model~IV describes a realistic scenario for a standard, truncated thin disk with an additional structure made by an ADAF-type corona which extends in the inner zones where the disk evaporates~\cite{Meyer-Hofmeister:2017ott, Meyer-Hofmeister:2012fol}. We therefore model the disk using the Shakura-Sunyev solution and the corona by the self-consistent solution described in~\cite{Narayan-Yi}. In our analysis, we vary the coronal density by several orders of magnitude to investigate its effect on the instability.
Furthermore, in thin disks the thickness is $H/R\ll1$. To try to capture this effect we multiply the radial Shakura-Sunyev profile by a $\sin^2 \theta$~\cite{Dima:2020rzg}. As a matter of fact, even more thinner profiles can be considered, but they would require higher angular resolution when computing the spectrum (see Sec.~\ref{sec:num}). As for the ADAF-type corona, the geometry is quasispherical so we can safely neglect deviations from spherical symmetry. Therefore, in Model~IV we consider the following effective mass: 

\begin{multline}
     \mu_{\rm eff, IV}^2(r,\theta)={\alpha}\left[\rho_H\Theta(r-r_0)\left(1-\sqrt{\frac{r_0}{r}}\right)^\frac{11}{20}\left(\frac{r_0}{r}\right)^{\frac{15}{8}}\times\right.\\
     \sin^2 \theta+\rho_C{\left(\frac{1}{r}\right)^{\frac{3}{2}}}\Big]\, .
\end{multline}

Finally, to explore the difference between the radial geometry of a thin and a thick disk, in Model~V we also consider a radial profile typical of a thick-disk axisymmetric model with the same corona as in Model~IV: 
\begin{multline}
     \mu_{\rm eff, V}^2(r,\theta)={\alpha}\left[\rho_H\Theta(r-r_0)\left(1-\frac{r_0}{r}\right)\left(\frac{r_0}{r}\right)^{\frac{3}{2}}\sin^2 \theta\right.\\+\rho_C{\left(\frac{1}{r}\right)^{\frac{3}{2}}}\Big]\,.
\end{multline}

Note that the salient features of these models can be qualitatively captured by three parameters, which, on physical grounds, should produce the following effects (see also Fig.~\ref{fig:confrontation}):  
\begin{itemize}
    \item Parameter $\rho_H$ represents the height of the barrier. If this value is high enough, it can naturally confine the scalar modes into a cavity. The higher the $\rho_H$ the more efficient the confinement. As $\rho_H$ represents a potential barrier rather than a bare mass (at variance with the standard superradiant instability from massive bosons), increasing $\rho_H$ should not stabilize the modes, but only confine them better.
    \item Parameter $r_0$ is the width of the cavity. If it is large enough, the barrier can efficiently confine the modes. In particular, a necessary requirement is that the width of the cavity must be greater than (or at least comparable to) the Compton wavelength of the modes~\cite{Cardoso:2004nk}.  In the following we consider two representative truncation values: $r_0=8M$ and $r_0=14M$. 
    \item Parameter $\rho_C$, instead, represents an offset that introduces an effective asymptotic mass to the scalar field, thus contributing to stabilizing the modes. Note also that, if the barrier is high enough and the modes are strongly confined in it, $\rho_C$ should be relevant only inside the cavity, because the part of the scalar field transmitted outside should be negligible. This effect will be explored by comparing Model~I with Model~II. 
\end{itemize}

In particular, as we shall later discuss, in the disk $\mu_{\rm eff}M\sim \sqrt{\alpha \rho_H} M$ should be sufficiently large for the barrier to confine the mode efficiently, whereas in the corona $\mu_{\rm eff}M\sim \sqrt{\alpha \rho_C} M$ corresponds to the gravitational coupling that governs the effective mass of the field inside the cavity. As such, $\sqrt{\alpha \rho_C} M\ll {\cal O}(0.1)$ for the instability not to be quenched.

\section{Numerical Methods for Non-Separable Equations in Arbitrarily Spinning Spacetime}
\label{sec:num}

In this section we present the numerical methods used to compute the spectrum of accreting spinning BHs in scalar-tensor theories.
We assume a stationary background and a $e^{-i \omega t}$ time dependence for the perturbation, where $\omega=\omega_R+i\omega_I$ is the (complex) eigenfrequency. Unstable modes correspond to solutions having $\omega_I>0$, which exponentially grow in time.
In the specific case of superradiant instabilities, this exponential growth is triggered if the mode satisfies the superradiant condition~\cite{Brito:2015oca}, i.e. $0<\omega_R<m \Omega_H=\frac{m a}{r_+^2+a^2}$, where $a M$ is the BH angular momentum, $r_+$ is the radius of its event horizon, and $m$ is the azimuthal number of the mode. 

We use a procedure consisting in two different numerical methods, both in the frequency domain. We first use a direct shooting method~\cite{Pani:2013pma} for finding solutions of Eq.~\eqref{eq:finalKGeq} in the case of spherical symmetry, i.e. for nonspinning BHs and when the effective mass profile depends only on the radial coordinate.
Imposing suitable boundary conditions at the horizon and at infinity, the shooting method allows us to solve the eigenvalue problem. Then, the wavefunctions and eigenfrequencies are used as starting guess solutions for computing the spinning case, by applying a numerical method suitable for nonseparable differential equations. In particular, following~\cite{Baumann:2019eav}, we express Eq.~\eqref{eq:finalKGeq} as a nonlinear eigenvalue problem which we solve with the nonlinear inverse iteration algorithm~\cite{guttel_tisseur_2017} (see below for details). Starting with the spherical symmetric case, we can iteratively solve the problem by gradually increasing the spin until we obtain the desired spinning configuration. With this method we can study also quasiextremal BHs and generic nonseparable equations. 

For the case of effective mass profiles having a $\theta$-dependence through $\sin^2\theta$, the field equations are nonseparable even for a nonspinning BH. In this case we introduce an extra iterative cycle in the procedure. We express the generic effective mass of any of the previous models as 
\begin{equation}
    \mu_{\rm eff}^2(r,\theta)=\mu_r^2(r)(1-k \cos^2\theta)+\mu_0^2(r)\, ,
\end{equation}
where we introduced the fictitious parameter $k$ connecting purely radial profiles ($k=0$) with $\theta$-depending profiles ($k=1$), whereas $\mu_0^2(r)$ comes from the BH corona. The extra cycle consists in applying the nonlinear inverse iteration to finding the mode of a nonspinning BH with a nonspherical density profile ($k=1$), using solutions with $k=0$ as starting guess: at each iteration we gradually increase $k$ and use the previous result as a guess, until we obtain the desired configuration with $k=1$ and zero BH spin. Finally, we use the latter solution as a starting guess to find the modes of a  spinning BHs with $k=1$, as previously explained.
Details of the numerical methods outlined below are given in the next subsections.

\subsubsection{Nonspinning BHs with radial density profiles: direct shooting method}

In the direct shooting method, the system is integrated from the horizon  to infinity. In particular, using the ansatz
\begin{equation}
    \varphi(t,r,\theta, \phi)=\sum_{l,m} \frac{R_{lm}(r)}{r}e^{-i\omega t}Y_{lm}(\theta, \phi)
\end{equation}
in spherical symmetry, the Klein-Gordon equation can be rearranged to obtain a Schrödinger-like equation
\begin{equation}
    \mathcal{D} R_{lm}=0
\end{equation}
where $f(r)=1-2M/r$, $M$ is the mass of the BH, and we defined the differential operator $\mathcal{D} \equiv   \frac{d^2}{dr_*^{2}}+ \omega^2- f(r) \Big(\frac{l(l+1)}{r^2}+\frac{2M}{r^3}+\mu_{\rm eff}^2\Big)$, where $r_*$ is the tortoise coordinate given by $dr/dr_*=f(r)$.  Owing to the spherical symmetry of the system, modes with different multipole numbers $l,m$ are decoupled. This equation is then solved by direct integration imposing suitable boundary conditions. In particular, at the horizon the solution must be a purely ingoing wave, given that the horizon behaves as a one-way membrane,
\begin{equation}
\label{eq:boundaryh}
    R_{lm}\sim e^{-i\omega r_*}\sum_n b_{n} (r-2M)^n\, ,
\end{equation}
while at infinity, the leading-order general solution reads:
\begin{equation}
R_{lm}\sim B e^{-k_{\infty}r_* }+C e^{+k_{\infty}r_*}\, ,
\end{equation}
where $k_{\infty}=\sqrt{\mu^2_\infty-\omega^2}$ and $\mu_\infty=\lim\limits_{r\to\infty}\mu_{\rm eff}(r,\theta)$. Usually, in the context of massive boson superradiant instabilities, the appropriate condition is $C=0$, implying exponentially damped solutions at infinity, i.e. quasibound states. Nevertheless, in our system the confinement is provided by a potential barrier in the vicinity of the BH, instead that by an asymptotic mass. In particular, in realistic accretion models the effective mass at infinity vanishes. The condition $C=0$ therefore would not correspond to damped solutions at infinity, but to ingoing waves from infinity. Clearly this solution is not physical, as it would correspond to an energy injection from infinity. Therefore, we must set the opposite conditions $B=0$, which is the one that corresponds to quasinormal modes~(QNMs), i.e. outgoing waves at infinity (see Ref.~\cite{Berti:2009kk} for a review). In some sense, we are hence looking for modes that were supposed to behave as QNMs if we did not have any effective mass; however, due to the scalar coupling to matter, these modes are confined by a barrier in the vicinity of the BH, and are thus prone to the superradiant instability if the BH spins sufficiently fast. 

We have also adapted a variation of the classical shooting method, where we integrate from the horizon to a fixed point and from infinity to the same point, and impose regularity of the wavefunction and its derivative to solve the equations~\cite{Pani:2013hpa}. We checked that the result is independent on the matching point and that the two methods give the same results.

\subsubsection{Nonseparable perturbations: Čebyšëv interpolation and nonlinear eigenvalue problem}

Let us now consider the case of nonseparable perturbations, which is relevant for both spinning BHs and even for nonspinning BHs if the effective mass depends on the angular coordinate $\theta$.

We assume an axisymmetric (Kerr) background so that perturbations have a definite azimuthal number $m$. We rewrite Eq.~\ref{eq:finalKGeq} in the following form:
\begin{align}\label{field_eq_baumann}
    &\nonumber\left\lbrace\displaystyle\frac{1}{\Delta(r)}[\boldsymbol{\mathcal{L}}^2+a^2\cos^2\theta(\mu_{\rm eff}^2(r,\theta)-\omega^2)]-\frac{1}{\Delta(r)}\frac{\partial}{\partial r}\left[\Delta(r)\frac{\partial}{\partial r}\right]\right.\\&\nonumber-\omega^2-\frac{P_+^2}{(r-r_+)^2}-\frac{P_-^2}{(r-r_-)^2}+\frac{A_+}{r-r_+}-\frac{A_-}{r-r_-}\\&+\left.\mu_{\rm eff}^2(r,\theta)\left(1+\frac{B_+}{r-r_+}-\frac{B_-}{r-r_-}\right)\right\rbrace\varphi(t,r,\theta,\phi)=0 \, ,
\end{align}
where $ A_\pm=\mp 2\omega^2 M+\frac{P_+^2+P_-^2 -(8 M^2-a^2)\omega^2}{r_+-r_-}$, $ B_\pm=\frac{2M^2-a^2}{r_+-r_-}\pm M$, $\lim_{r\to\infty}\mu_r(r)=0$, $ r_\pm=M\pm\sqrt{M^2-a^2}$, $ P_\pm=\frac{m a-2\omega M r_\pm}{r_+-r_-}$, $\boldsymbol{\mathcal{L}}^2=-\frac{1}{\sin\theta}\frac{\partial}{\partial\theta}\left(\sin\theta\frac{\partial}{\partial\theta}\right)-\frac{1}{\sin^2\theta}\frac{\partial^2}{\partial\phi^2}$, and $\Delta(r)=(r-r_+)(r-r_-)$. 
Note that the dependence on $k$ is contained inside $\mu^2_{\rm eff}(r,\theta)$ in the above equation.

At the horizon we must have ingoing waves,
\begin{equation}
    \displaystyle\varphi\sim (r-r_+)^{ i P_+}\,,
\end{equation}
whereas, as previously discussed, we impose that there are no waves coming from infinity, 
\begin{equation}
    \varphi\sim r^{-1 - \frac{ M \left(2\omega^2-\mu^2_\infty\right)}{k_\infty}} e^{k_\infty r}\,.
\end{equation} 
We apply the following ansatz for the scalar field~\cite{Baumann:2019eav}: 
\begin{equation}\label{phi_ansatz}
\varphi(t,r,\theta,\phi)= F(r)\sum\limits_{l,m}B_{lm}(\zeta(r))Y_{lm}(\theta,\phi)e^{-i \omega t}\,,
\end{equation}
where 
\begin{multline}
\small F(r)=\left(\frac{r-r_+}{r-r_-} \right)^{i P_+} (r-r_-)^{-1 - \frac{ M \left(2\omega^2-\mu^2_\infty\right)}{k_\infty}}e^{k_\infty (r-r_+)}
\end{multline}
captures the asymptotic behaviors of the solution.
Henceforth for simplicity we drop the index $m$ from $B_{lm}$. In the numerical results presented in the next section we will always consider the case $m=1$. In the above ansatz $B_l(\zeta(r))$ are radial functions depending on the auxiliary radial coordinate $\zeta\in(-1,1)$, defined by the following mapping
\begin{equation}\label{mapping}
    \zeta(r)=\frac{r-\sqrt{4 r_+ (r-r_-)+r_-^2}}{r-r_-}\, ,
\end{equation}
\begin{equation}
    r(\zeta)=\frac{4 r_++r_- (\zeta^2-1)}{(\zeta-1)^2}\,.
\end{equation}
By performing a spherical harmonics decomposition of Eq.~\ref{field_eq_baumann}, we obtain an infinite cascade of coupled radial equations:
\begin{multline}\label{radial_eqs}
    \left[\frac{\partial^2}{\partial\zeta^2}+C^{(1)}_l(\zeta)\frac{\partial}{\partial\zeta}+C^{(2)}_l(\zeta) \right]B_l(\zeta)\\+\sum\limits_{j=-4}^4 C^{(3)}_{l,j}(\zeta)B_{j}(\zeta)=0\, ,
\end{multline}
where we have the following expressions for the couplings
\begin{multline}
    C^{(3)}_{l,j}(\zeta)=-\frac{c^{(1)}_{l,j}}{\zeta'^2(r(\zeta))}\left\lbrace\frac{a^2[\mu^2_r(r(\zeta))+\mu^2_0(r(\zeta))-\omega^2]}{\Delta(r(\zeta))}\right.\\\left.-k \mu_r^2(r(\zeta))\left[1+\frac{B_+}{r(\zeta)-r_+}-\frac{B_-}{r(\zeta)-r_-}\right]\right\rbrace\\+\frac{k c^{(2)}_{l,j}a^2 \mu_r^2(r(\zeta))}{\zeta'^2(r(\zeta))\Delta(r(\zeta))}\, ,
\end{multline}
with the Clebsch-Gordan coefficients:
\\\\
$\displaystyle c^{(1)}_{l,j}=\left\langle l,m\right|\cos^2\theta\left|j,m\right\rangle=$
\begin{equation}
    ~=\frac{1}{3}\delta_{lj}+\frac{2}{3}\sqrt{\frac{2j+1}{2l+1}}\left\langle j,2,m,0\right|\left.l,m\right\rangle\left\langle j,2,0,0\right|\left.l,0\right\rangle\,,
\end{equation}
$\displaystyle c^{(2)}_{l,j}=\left\langle l,m\right|\cos^4\theta\left|j,m\right\rangle=$
\begin{multline}
    ~=\frac{1}{5}\delta_{lj}+\frac{4}{7}\sqrt{\frac{2j+1}{2l+1}}\left\langle j,2,m,0\right|\left.l,m\right\rangle\left\langle j,2,0,0\right|\left.l,0\right\rangle\\+\frac{8}{35}\sqrt{\frac{2j+1}{2l+1}}\left\langle j,4,m,0\right|\left.l,m\right\rangle\left\langle j,4,0,0\right|\left.l,0\right\rangle\,,
\end{multline}
and the following expressions for the remaining functions
\begin{multline}
C^{(1)}_l(\zeta)=\left(\frac{1}{r(\zeta)-r_+}+\frac{1}{r(\zeta)-r_-} \right)\frac{1}{\zeta'(r(\zeta))}+\\\frac{1}{\zeta'(r(\zeta))}\frac{2 F'(r(\zeta))}{F(r(\zeta))}+\frac{\zeta''(r(\zeta))}{\zeta'^2(r(\zeta))}\,,
\end{multline}
\begin{multline}
C^{(2)}_l(\zeta)=\frac{1}{\zeta'^2(r(\zeta))}\left\lbrace\frac{ F''(r(\zeta))}{F(r(\zeta))}+\left[\frac{1}{r(\zeta)-r_+}+\right.\right.\\\left.\frac{1}{r(\zeta)-r_-} \right]\frac{F'(r(\zeta))}{F(r(\zeta))}+\frac{P_+^2}{[r(\zeta)-r_+]^2}+\frac{P_-^2}{[r(\zeta)-r_-]^2}\\-\left[\mu_r^2(r(\zeta))+\mu_0^2(r(\zeta))\right]\left[1+\frac{B_+}{r(\zeta)-r_+}-\frac{B_-}{r(\zeta)-r_-}\right]\\-\left.\frac{A_+}{r(\zeta)-r_+}+\frac{A_-}{r(\zeta)-r_-}+\omega^2-\frac{l(l+1)}{\Delta(r(\zeta))}\right\rbrace \, .
\end{multline}

The couplings $c^{(1)}_{l,j}$ are nonzero for $j\in\left\lbrace l,l\pm 2 \right\rbrace$, while $c^{(2)}_{l,j}$ are nonzero for $j\in\left\lbrace l,l\pm 2, l\pm 4 \right\rbrace$, thus each $l$-mode is coupled with 4 other differing ones. In order to find solutions we truncate the infinite tower to some $L$ (i.e. we neglect perturbations with $l\leq L$) and transform the remaining (finite) set of radial equations into a matrix form. The radial coordinate is then discretized through a Čebyšëv interpolation, which is defined by the following polynomials
\begin{align}
    p_n(\zeta)&=\displaystyle\frac{\prod_{q\neq n}(\zeta-\zeta_q)}{\prod_{q\neq n}(\zeta_n-\zeta_q)}=\frac{p(\zeta)w_n}{\zeta-\zeta_n}\,,\\
    p(\zeta)&=\displaystyle\prod\limits_{q=0}^N(\zeta-\zeta_q)\,,
\end{align}
with Čebyšëv nodes
\begin{equation}
    \zeta_n=\cos\left(\frac{\pi(2n+1)}{2(N+1)}\right)\,,
\end{equation}
and corresponding weights~\cite{Baumann:2019eav,lagrange_interpol_1,lagrange_interpol_2}
\begin{equation}
    w_n=\frac{1}{p'(\zeta_n)}=(-1)^n\sin\left(\frac{\pi(2n+1)}{2(N+1)}\right)\,.
\end{equation}
where $N+1$ is the number of interpolation points and $n\in\left[0,N\right]$.
The radial functions $B_l$ are hence described by a set of $(L+1)(N+1)$ coefficients $B_l(\zeta_k)$, that define a $(L+1)(N+1)$-dimensional array $\underline{B}$, while the radial equations take the form
\begin{multline}
\sum\limits_{q=0}^N \left[ p_q''(\zeta_n)B_l(\zeta_q)+ C^{(1)}_l(\zeta_n)p_q'(\zeta_n)B_l(\zeta_q)\right]+\\ C^{(2)}_l(\zeta_n)B_l(\zeta_n)+\sum\limits_{j=-4}^4 C^{(3)}_{l,j}(\zeta_n)B_{j}(\zeta_n)=0
\end{multline}
By exploiting the second barycentric form of the Lagrange polynomials, we can get numerically robust differentiation matrices~\cite{Baumann:2019eav,lagrange_interpol_1,lagrange_interpol_2}:
\begin{equation}
 p_q'(\zeta_n)=\left\lbrace \begin{matrix}
~~~~~~\frac{w_q/w_n}{\zeta_n-\zeta_q}~~~~~~~~~~~n\neq q~~~~~~~~~~~~~~~~~~\\~\\
-\sum\limits_{b,b\neq n}^N p_b'(\zeta_n)~~~~~n=q~~~~~~~~~~~~~~~~~~
\end{matrix} \right.
\end{equation}
\begin{equation}
p_q''(\zeta_n)=\left\lbrace \begin{matrix}
~~~~2 p_q'(\zeta_n)\left(p_n'(\zeta_n)-\frac{1}{\zeta_n-\zeta_q} \right)~~~~~~n\neq q\\~\\
2 p_q'(\zeta_n)p_n'(\zeta_n)+\sum\limits_{b,b\neq n}^N \frac{2 p_b'(\zeta_n)}{\zeta_n-\zeta_b}~~~~n=q
\end{matrix} \right.
\end{equation}
At the end of this procedure we obtain a nonlinear eigenvalue problem in $\omega$ and $\underline{B}$,
\begin{equation}
    \textbf{A}(\omega)\underline{B}=0\,,
\end{equation}
to be solved through nonlinear inverse iteration~\cite{guttel_tisseur_2017}.

\section{Results}\label{sec:results}

\begin{figure*}[t]
\centering
\includegraphics[width=0.46\textwidth]{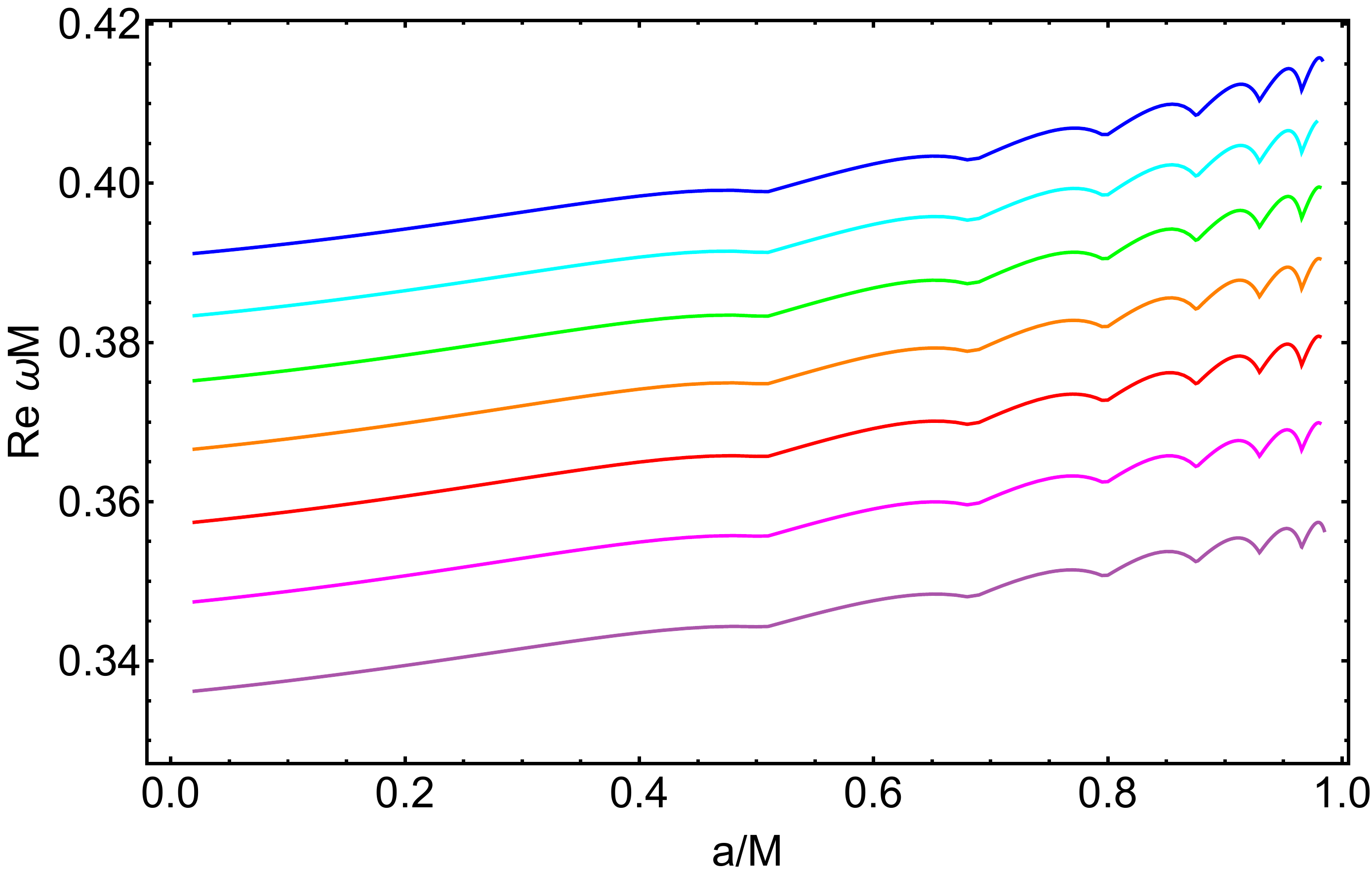}
\includegraphics[width=0.49\textwidth]{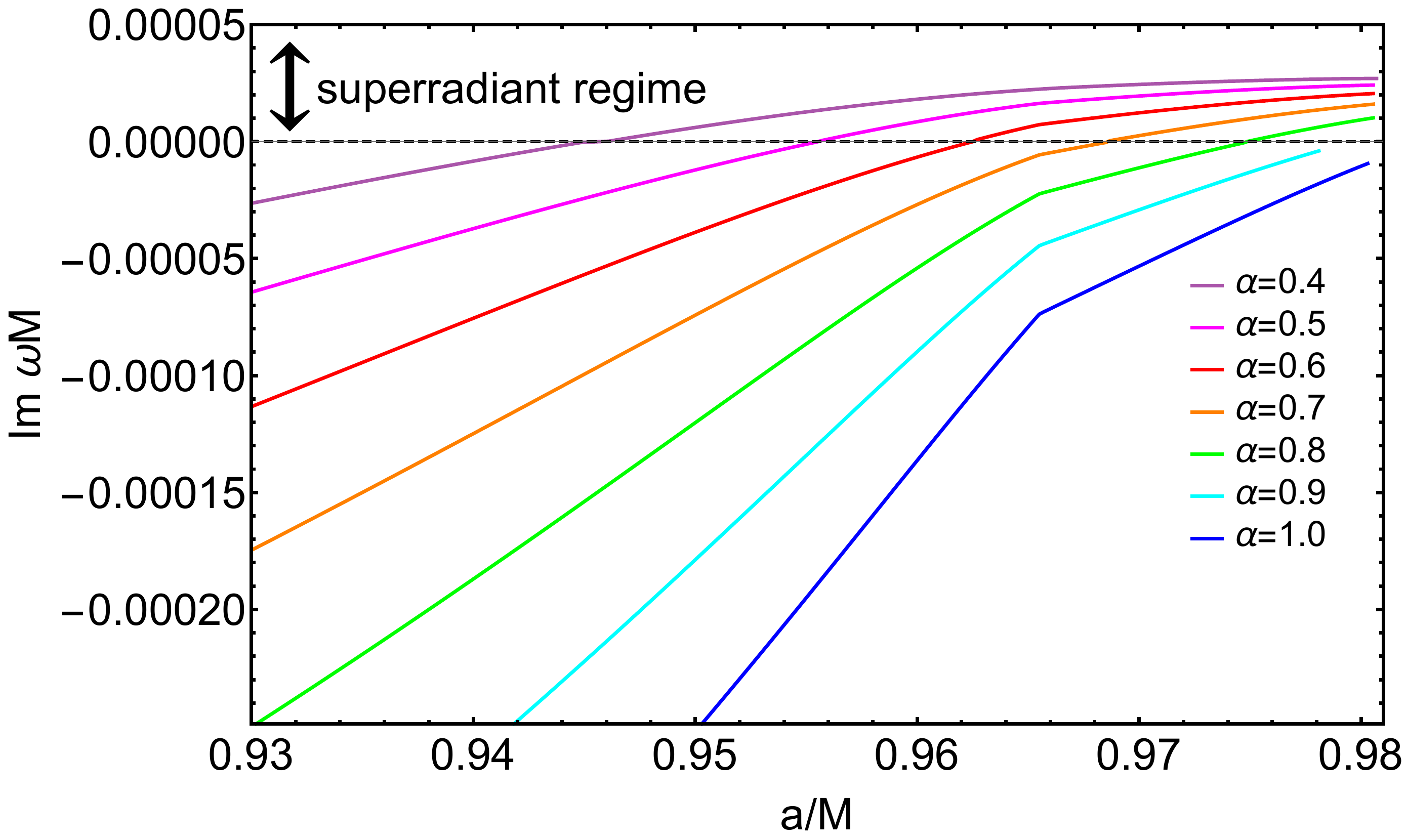}
\caption{Real (left panel) and imaginary (right panel) part of the modes in Model~I as a function of the BH spin for different values of the coupling ${\alpha}$. For lower values of this parameter, $\omega_R$ decreases and the modes become superradiantly unstable for smaller values of the BH spin.}
\label{fig:Dima}
\end{figure*}

\subsection{Models I: key ingredients for the instability}
We start by studying the first three models with the same density profiles considered in~\cite{Dima:2020rzg}, to show that the obstacles existing in plasma-driven superradiant instabilities can be circumvented in scalar-tensor theories. 
Figure~\ref{fig:Dima} shows the modes of Model~I with $\rho_H=4/M^2,\rho_C=0.09/M^2$, $r_0=8M$, and different values of ${\alpha}$. For ${\alpha}=1$ we recover the results obtained in~\cite{Dima:2020rzg}. In this case, superradiance does not appear before $a/M=0.99$. However, if we consider lower values of ${\alpha}$ the effective mass of the scalar field (and hence the superradiant mode frequency) decreases and the superradiant condition is fulfilled for smaller values of the spin. This is evident by looking at the real part of the mode in the left panel of Fig.~\ref{fig:Dima}. As the coupling ${\alpha}$ decreases, the real part becomes smaller, eventually entering the superradiance condition. Therefore, while in plasma-driven superradiant instabilities in GR a small increase of the coronal mass is sufficient to quench the instability~\cite{Dima:2020rzg}, in scalar-tensor theories decreasing ${\alpha}$ is sufficient to circumvent this obstacle and recover an efficient superradiant regime, as also discussed more in detail below.

Nevertheless, by decreasing ${\alpha}$ too much, the potential barrier becomes too low and is not able to confine the modes. For the case of Model~I, we numerically find that when ${\alpha}<0.15$ the eigenfunctions start having a nonnegligible amplitude even after the potential barrier, suggesting that the confinement starts becoming inefficient.

\begin{figure}[t]
\centering
\includegraphics[width=0.49\textwidth]{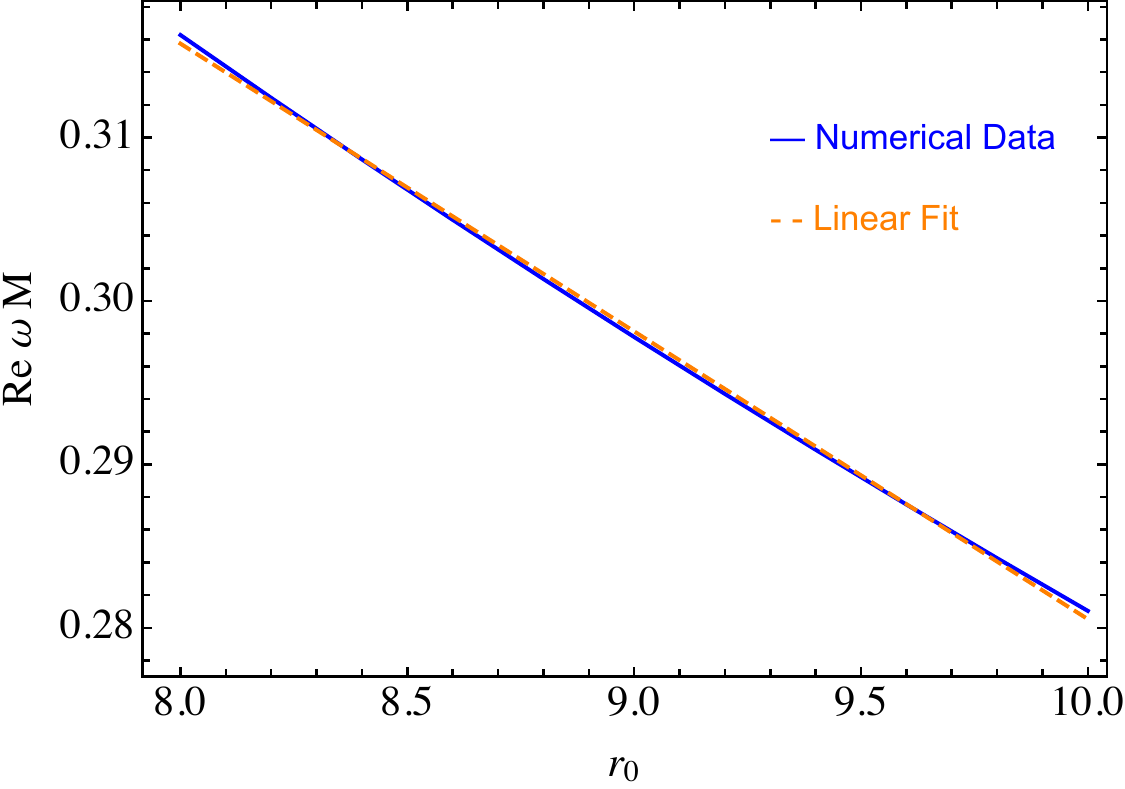}
\caption{Real part of the modes in Model~I as a function of $r_0$ for $\alpha \rho_H M^2=4$, $\alpha \rho_C M^2=0$, and $a=0$. The real part decreases linearly with $1/r_0$, as can be observed by comparing the numerical result with a linear fit.}
\label{fig:BHBomb}
\end{figure}

Assuming a high spinning BH, the superradiant instability can therefore be quenched in the following cases:
\begin{itemize}
\item{If the density of the corona is high enough to stabilize the system. In Model~I and for the chosen parameters, this happens when $\sqrt{\alpha \rho_C}M>0.42$.}
\item{If the barrier is not high enough to confine modes. This starts happening when $\sqrt{\alpha \rho_H}M<0.76$.}
\item{If the width of the cavity is not sufficiently large as to support quasibound states inside it. Indeed, when the effective mass within the cavity is negligible (i.e., $\sqrt{\alpha \rho_C}M\ll0.1$), this system resembles the original BH bomb, where the frequencies scale as the inverse of the width of the cavity, $\omega_R\sim 1/r_0$~\cite{Cardoso:2004nk}. In Fig.~\ref{fig:BHBomb} we show that we recover the same scaling in our system.}
\end{itemize}

Reversing the argument, if the barrier is high enough and the cavity wide enough, modes can be confined efficiently. If in addition the coronal density is tenuous enough not to provide modes in the cavity with a too large effective mass, then an efficient superradiant instability can develop around an accreting spinning BH. We shall come back to this point in Sec.~\ref{sec:contraints}.

For the time being we wish to stress that the main difference with respect to~\cite{Dima:2020rzg} is the free parameter ${\alpha}$ appearing in scalar-tensor theories. In~\cite{Dima:2020rzg}, it was shown that, even though the disk can create a cavity where superradiant modes can develop, an extremely tenuous plasma inside this cavity (of the order of $n_e\sim 10^{-2} {\rm cm}^{-3}$ for $M=10M_\odot$) is sufficient to quench the instability. Given that realistic coronal densities are orders of magnitude higher, the instability is strongly suppressed.
On the other hand, as discussed in detail in Sec.~\ref{sec:contraints} below, in our system
there are large unconstrained ranges of $\alpha$ in which the effective mass due to the corona is negligible and, \emph{yet}, the disk barrier is sufficiently high.

\subsection{Models~II and III: truncation of the corona and smoothness of the profiles}
Model~II aims to quantitatively verify that only the coronal density \textit{inside} the cavity is relevant for providing an additional effective mass. For this reason, we truncate the corona at $r_0$, where the disk begins. We obtain numerical results which almost coincide with those of Model~I, confirming that what is really relevant to increase the effective mass --~and hence to possibly quench the instability~-- is only the density inside the cavity. 

\begin{figure}[t]
\centering
\includegraphics[width=0.49\textwidth]{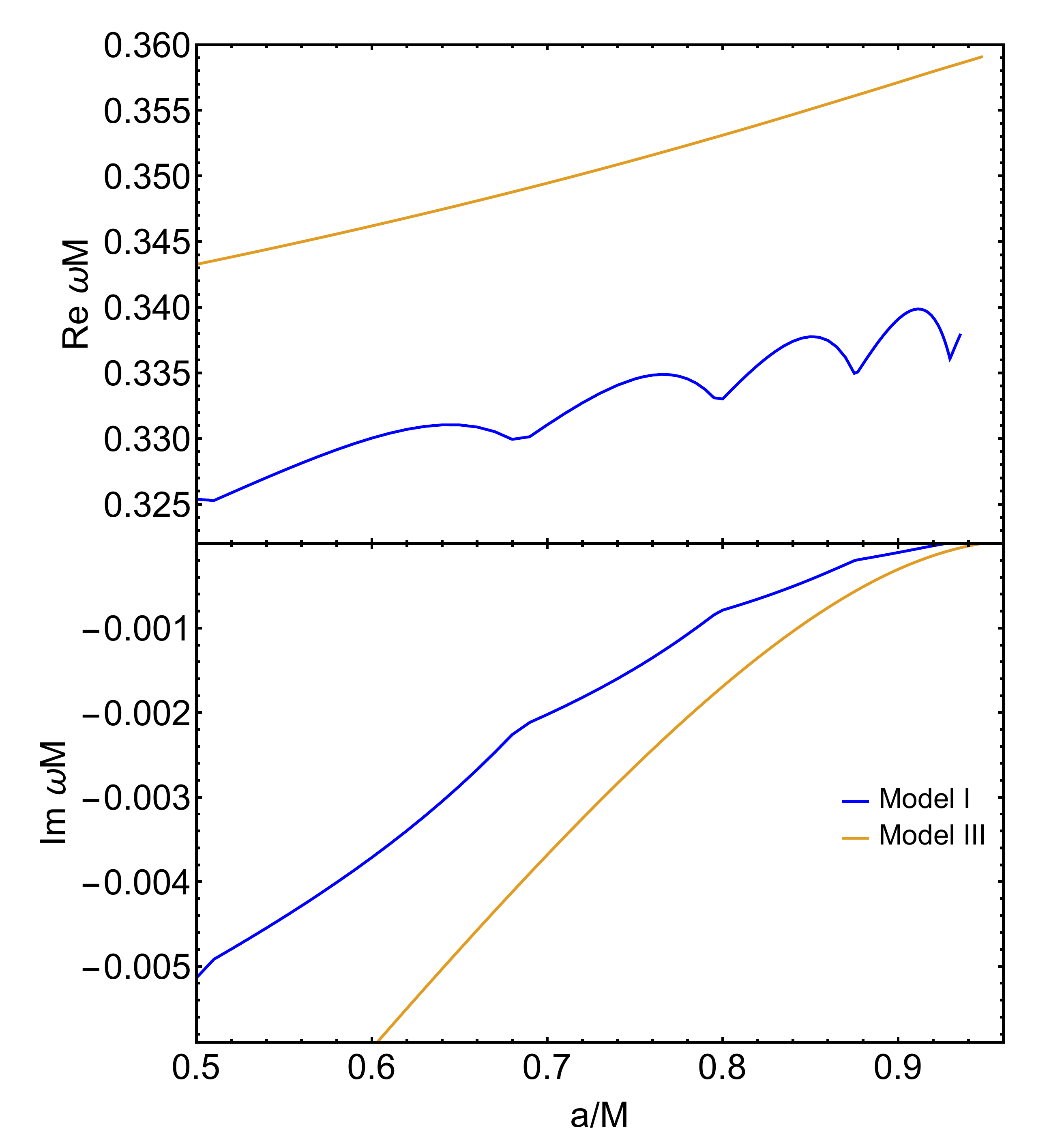}
\caption{Real and imaginary parts of modes respectively from Model~I (blue) and Model~III (orange). By replacing the step function with a sigmoid, the profile becomes more regular and the corners disappear.}
\label{fig:ComparisonModes}
\end{figure}

Finally, in Model~III we replace the step function of the inner edge by a sigmoid, in order to show that the corners in both the real and imaginary parts shown in Fig.~\ref{fig:Dima}  are an artifact of the Heaviside function used in modelling the density profile. In Fig.~\ref{fig:ComparisonModes} we show that when the barrier is instead described by a smooth sigmoid, the corners disappear, and the resulting modes are also smooth functions of the model parameters.

\subsection{Models~IV and V: role of the corona density}
\begin{figure*}[t]
\centering
\includegraphics[height=0.49\textwidth]{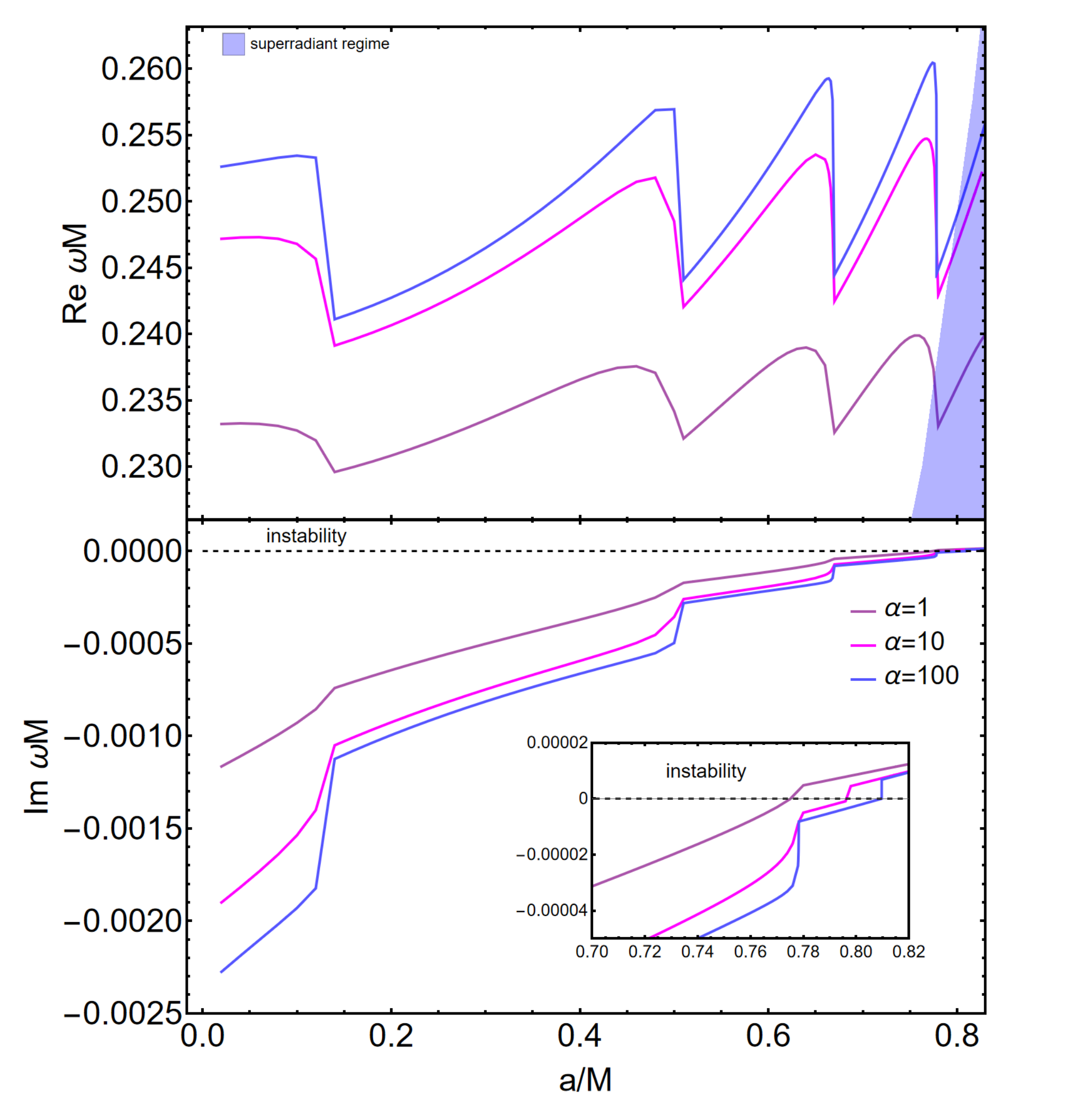}
\includegraphics[height=0.49\textwidth]{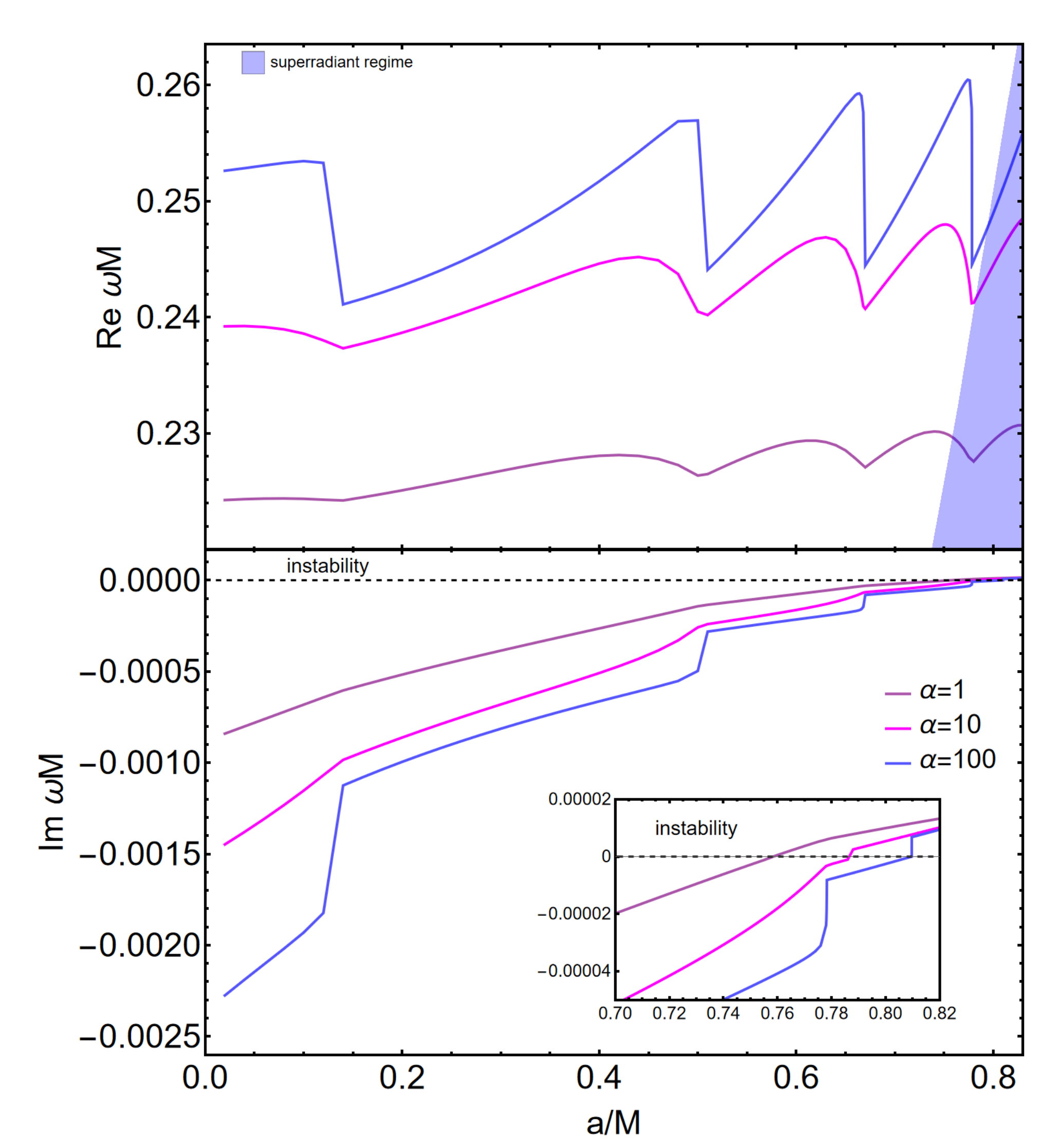}
\caption{Superradiant modes of Model~IV (left) and Model~V (right) for $r_0=14M$ and $\gamma=10^{-6}$ as functions of the dimensionless spin parameter for different values of ${\alpha}$. Even by varying ${\alpha}$ across two orders of magnitude, the instability is preserved.}
\label{fig:realistic}
\end{figure*}

In these models, we study the impact of different coronal density by parametrizing $\rho_C=\gamma \rho_H$ and varying the parameter $\gamma$ in the realistic range $10^{-6}-10^{-1}$ (see e.g.~\cite{Bisnovatyi-Kogan:1976fbc, DeVilliers:2003gr, Meyer-Hofmeister:2017ott}). Figure~\ref{fig:realistic}  show the imaginary part of the solutions for $\gamma=10^{-6}$, $r_0=14M$, $\rho_H=4/M^2$ obtained by varying the parameter ${\alpha}$ in Model~IV and Model~V. By varying ${\alpha}$ across two orders of magnitude the instability is preserved with qualitatively similar features: this is because the coronal density is so low that it remains negligible, while the disk density is sufficiently high to confine the modes in this range of $\alpha$. Thus, if the coronal density is strongly suppressed with respect to the disk one, it is possible to have an instability in a wide range of the coupling ${\alpha}$. Also note that assuming a larger truncation radius yields a smaller spin threshold for the instability. This is because, akin to the original BH bomb phenomenon, the real part of the frequency decreases with the truncation radius $\omega_R \sim 1/r_0$~\cite{Cardoso:2004nk} (see Fig.~\ref{fig:BHBomb}). 

\begin{figure}[t]
\centering
\includegraphics[width=0.49\textwidth]{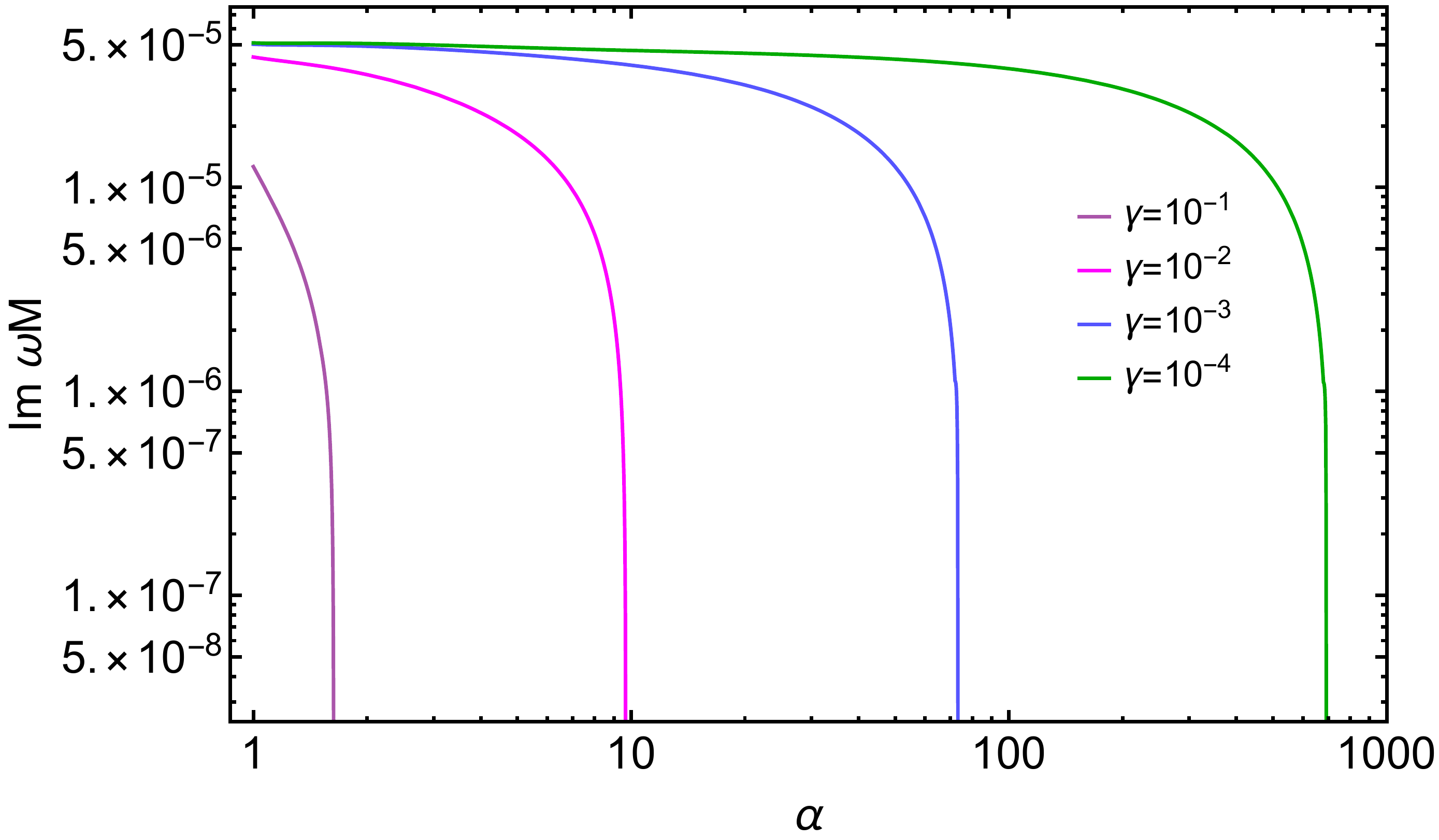}
\caption{Imaginary part as a function of $\alpha$ in Model~V for different values of the density ratio $\gamma$ between the corona and the disk for a spinning BH with $a=0.97M$. When $\alpha\gamma\gtrsim O(10^{-1})$, the instability is suppressed. Hence, the lower $\gamma$, the more efficient the instability is across several orders of magnitudes in $\alpha$.}
\label{fig:realistic_alpha1}
\end{figure}

Finally, Fig.~\ref{fig:realistic_alpha1} shows the imaginary part of the modes as a function of $\alpha$ for different density ratios $\gamma$ in Model~V with $\rho_H=4/M^2$ and $r_0=8M$. Note that, for certain values of $\alpha$ (e.g. $\alpha\approx 1$ for the parameters chosen in Fig.~\ref{fig:realistic_alpha1}) the modes are independent of $\gamma$ in the $\gamma\ll1$ limit. This is because the coronal density in this regime is subdominant and does not affect the mode. On the other hand, as the $\alpha$ parameter grows, the coronal effective mass eventually becomes relevant and quenches the instability. In particular, for the chosen parameters the instability is suppressed when $\alpha\gamma\gtrsim O(10^{-1})$.  

\section{Constraints on scalar-tensor theories from spinning BH observations}
\label{sec:contraints}
After having explored the parameter space of our models and having identified the key features of the plasma-triggered superradiant instability in scalar-tensor theories, we are now in a position to draw a general picture and use it to identify the parameter space of scalar-tensor theories in which the instability is effective.

The first key ingredient is a sufficiently dense disk that extends down to the BH up to some truncation radius $r_0>{\cal O}({\rm few})M$, as predicted in various models. The requirement that the disk can effectively confine scalar modes implies 
\begin{equation}
    \sqrt{\alpha \rho_H}M\gtrsim 1\,.\label{cond1}
\end{equation}

For a standard thin disk the typical outer density is~\cite{1973A&A....24..337S, env2014}: 
\begin{equation}
    \rho \approx 169 \frac{f_{\rm Edd}^\frac{11}{20}}{(r/M)^\frac{15}{8}}\Bigg(1-\sqrt{\frac{r_0}{r}}\Bigg)^{\frac{11}{20}}\Bigg(\frac{0.1}{\beta}\Bigg)^{\frac{7}{10}}M_6^{-\frac{7}{10}}{\rm kg/m}^3 \, ,
\end{equation}
where $r_0$ is the truncation radius, $\beta$ is the viscosity parameter, $f_{\rm Edd}=\Dot{M}/\Dot{M}_{\rm Edd}$ is the mass accretion Eddington ratio, and we defined $M_6=M/(10^6 M_\odot)$.
Using the above normalization, Eq.~\eqref{cond1} yields a \emph{lower} bound on the scalar coupling, 
\begin{equation}
   \alpha \gtrsim \alpha_c=\frac{1}{\rho_H M^2}\approx 3\times 10^6 M_6^{-13/10}\,,
\end{equation}
so that supermassive BHs would yield a smaller lower bound.

The above condition is necessary but not sufficient. In the presence of a corona with characteristic density $\rho_C=\gamma \rho_H$, one should also require that the effective mass inside the cavity be not too large, namely,
\begin{equation}
    \sqrt{\alpha \rho_C}M\lesssim1\,. \label{cond2}
\end{equation}
This condition can be written as a \emph{upper} bound on the scalar coupling,
\begin{equation}
    \alpha\lesssim \frac{\alpha_c}{\gamma}\approx \frac{3}{\gamma}\times 10^6 M_6^{-13/10}\,.
\end{equation}
Since the corona is much less dense than the disk, $\gamma\ll1$ and condition~\ref{cond1} has always some overlap with condition~\ref{cond2}. In particular, provided the disk truncation is not too close to the BH horizon, the superradiant instability can occur when
\begin{equation}
  3\times 10^6 \lesssim \alpha M_6^{13/10}\lesssim 3 \left(\frac{10^{-4}}{\gamma}\right) 10^{10}\,, \label{condF}
\end{equation}
where we have normalized the typical coronal density such that $\gamma=\rho_C/\rho_H=10^{-4}$.

Remarkably, different classes of BHs could constrain different ranges of $\alpha$, extending roughly from $\alpha\sim {\cal O}(100)$ for $M\sim 10^9 M_\odot$ up to $\alpha\sim {\cal O}(10^{17})$ for $M\sim 5 M_\odot$. 
Furthermore, as shown in the previous section the instability time scale, $\tau=1/\omega_I$, is typically very short compared to astrophysical time scales. The instability can therefore be effective to change the dynamics of the system (see~\cite{Brito:2014wla,Brito:2015oca} for the phenomenology of the BH superradiant instability in various systems).

This implies that, providing the accretion flow can be accurately modelled, constraints on scalar-tensor theories coming from the observation of highly-spinning accreting BHs can rule out scalar-tensor theories with positive couplings in a very wide range. Interestingly, while there exists stringent constraints on $\alpha<0$ coming from spontaneous scalarization and the absence of dipolar radiation in binary pulsars~\cite{Cardoso:2013opa, Cardoso:2013fwa,Kramer:2021jcw}, the regime where $\alpha>0$ is essentially unconstrained and is relevant for cosmology. 

The $\alpha\gg1$ regime is particularly interesting for certain scalar-tensor theories. For example, in the symmetron model~\cite{Hinterbichler:2010es} the conformal factor reads\footnote{The bare mass term and scalar self-interactions of this cosmological model are negligible for astrophysical BHs~\cite{Davis:2011pj,Davis:2014tea}, so the approximations assumed in Sec.~\ref{sec:setup} apply.} $A(\phi)=1+{\alpha} \phi^2/2$ and requiring the Milky Way to be screened imposes ${\alpha} \gtrsim 10^6-10^8$~\cite{Hinterbichler:2011ca, Davis:2014tea, deAguiar:2021bzg}, which perfectly lies in the range that can be potentially excluded by accretion-driven BH superradiance.

\section{On the role of nonlinearities for plasma-driven superradiant instability in scalar-tensor theories}\label{sec:nonlinear}

As previously discussed, we find a wide range of parameter space prone to trigger matter-driven BH superradiant instabilities in scalar-tensor theories. Since during the instability the amplitude of the scalar field grows exponentially in a short timescale, linear theory eventually breaks down. It is therefore crucial to understand the modifications that nonlinearities will introduce in the system.  This can be done by analysing the backreaction of the superradiantly growing scalar field on to the plasma. In the Jordan frame, plasma particles follow geodesics, as it can be easily seen by the conservation of the matter stress energy tensor:
\begin{equation}
    \nabla_\nu T^{\mu\nu}=0 \rightarrow \frac{D u^\mu}{D\tau}=u^\nu \nabla_\nu u^\mu=0,
\end{equation}
where $u^\mu$ is the plasma four velocity in the Jordan frame.
Switching to the Einstein frame, this equation can be rewritten as (see e.g.~\cite{Fujii:2003pa}):
\begin{equation}
    \frac{D u_E^\mu}{D\tau_E}=f_\nu u_E^\nu u_E^\mu-f_E^\mu (u_{E\,\mu} {u_E}^\mu),
\end{equation}
where $u_E^\mu=d x^\mu /d\tau_E$ and $\tau_E$ are the four velocity and proper time in the Einstein frame, respectively, whereas $f_\nu=-\partial_\nu \ln A(\Phi)$ and $f_E^\mu=g_E^{\mu\nu}f_{\nu}$. By expanding the conformal factor around $\Phi\sim \Phi^{(0)}$ as before, this equation can be rewritten to the leading order as
\begin{equation}
    \frac{D u_E^\mu}{D\tau_E}= - \alpha\Big(\varphi \partial_\nu \varphi u_E^\mu u_E^\mu - g_E^{\mu\alpha}\varphi \partial_\alpha \varphi ({u_E}^\nu u_{E\,\nu})\Big).
\end{equation}
From this equation it is possible to observe that the acceleration of the plasma particles in the Einstein frame depends on nonlinear terms in the scalar field $\varphi$, with coupling constant $\alpha$. By solving this equation it is then possible to relate the backreaction on the four velocity with the backreaction on the density via the continuity equation of the fluid. Hence, nonlinear effects can modify the density of the fluid, which evolves dynamically. The details on the evolution depend on the specific models and on higher-order scalar interactions in the scalar-tensor theories.

Nevertheless, and most crucially, this system is safe from another nonlinear effect, the relativistic transparency, which severely hampers plasma-driven superradiant instabilities in GR~\cite{Cardoso:2020nst}. Due to this nonlinear correction, the effective photon mass in a plasma is modified in the relativistic regime~\cite{1970PhFl...13..472K,1971PhRvL..27.1342M,Cardoso:2020nst}:
\begin{equation}
    \omega_p^2=\frac{4\pi e^2 n}{m_e\sqrt{1+\frac{e^2 E^2}{m_e^2 \omega^2}}}.
\end{equation}
In the presence of large-amplitude electric fields, the effective mass vanishes, which dramatically quenches plasma-driven GR instabilities before a significant amount of energy can be extracted from the BH~\cite{Cardoso:2020nst}. This effect can be interpreted as a relativistic increase of the relativistic electron mass-energy, and it is therefore a completely different effect from the field backreaction on the density distribution. We will now show that in scalar-tensor theories the effective mass does not suffer from a similar suppression. Indeed, in this system the effective mass is the trace of the stress-energy tensor, $T^{\mu\nu}=\rho u^\mu u^\nu$.
The crucial point is that, no matter what the fluid four-velocity is, the trace of this tensor is always the rest-mass density, given that $u_\mu u^\mu=-1$ is a relativistic invariant. Therefore, even if the plasma is accelerated to relativistic velocities, the expression of the effective mass does not change (although the density becomes a dynamical quantity as discussed before). This follows from the fact that the trace of a tensor is a scalar quantity, which is invariant under Lorentz boosts. Hence, no Lorentz boost factor enters in the effective scalar mass in the relativistic nonlinear regime, at variance with the standard case of plasma-photon interactions.

\section{Conclusion and extensions}
\label{sec:discussion}
We have studied in detail the phenomenon of matter-driven BH superradiant instabilities in scalar-tensor theories. We have considered arbitrarily spinning BHs and realistic models of truncated thin and thick accretion disks.
In general the linearized scalar equation is nonseparable, and we have discussed in detail an efficient numerical method to find the unstable modes for this system.

We found two interesting results: i) although the qualitative features of the instability are akin to the case of plasma-driven electromagnetic superradiant instabilities within GR, the obstacles preventing the latter (namely suppression due to the corona~\cite{Dima:2020rzg} and nonlinearities~\cite{Cardoso:2020nst}) can be circumvented in scalar-tensor theories; ii) Remarkably, there exists a very wide range of (positive and large) scalar couplings where BH superradiant instabilities can be triggered in realistic scenarios. This range is unconstrained by observations and it actually includes the regime where certain scalar-tensor alternatives to the dark energy, e.g. symmetron models with screening, can evade solar system constraints while remaining cosmologically viable.
Our results suggest that such theories could be ruled out as dark-energy alternatives by the observation of highly spinning BHs, using the same technique adopted to constrain ultralight bosons from BH mass-spin observations~\cite{Arvanitaki:2009fg,Arvanitaki:2010sy,Brito:2015oca}.
However, at variance with the ultralight boson case, here an accurate modelling of the accretion flow around the BH is needed in order to quantitatively characterize the instability.

Furthermore, the possibility of circumventing nonlinear damping effects suggests that the models proposed for ordinary plasma-driven instabilities (e.g. as a possible explanation for fast radio bursts~\cite{Conlon:2017hhi} or for constraints on primordial BHs~\cite{Pani:2013hpa}) could actually work in the context of scalar-tensor theories.

Although the quantitative features of the instability depend on the geometry of the accretion flow near a BH, the key ingredients are naturally predicted in various models: i)~a sufficiently dense disk with a sharp transition from a low-density to a high-density region in the vicinity of the ISCO; ii)~A sufficiently tenuous corona in the low-density region, such that its density is much smaller than the one of the disk; iii)~a BH spinning sufficiently fast to make the quasibound modes unstable against the superradiant instability.

The numerical method implemented to compute the unstable modes in the absence of separable equations is general and robust, and could find applications in other contexts.

Another interesting finding is the fact that the unstable modes of this system resemble a quasibound state in the vicinity of the BH but are in fact propagating waves far from it. Therefore, one could imagine situations in which (perhaps during the superradiant growth) the quasibound states are not efficiently trapped and could propagate to infinity, possibly after several reflections within the cavity.
The scalar modes in the Einstein frame correspond to a (breathing) scalar polarization of the gravitational waves in the Jordan frame. Therefore, the phenomenology of this effect would be similar to the gravitational-wave echoes predicted for matter fields~\cite{Barausse:2014pra}, near-horizon structures~\cite{Cardoso:2016rao}, and exotic compact objects~\cite{Cardoso:2019rvt}. A more detailed study of this interesting phenomenon, that we leave to the future, will probably require a time-domain analysis. 

Finally, an important follow-up of our work is to study backreaction effects on the plasma and the full dynamics of the system at the nonlinear leve.

\begin{acknowledgments}

We thank Vitor Cardoso for comments on the manuscript and Riccardo La Placa for useful conversations about accretion physics.
We acknowledge the financial support provided under the European Union's H2020 ERC, Starting Grant agreement no.~DarkGRA--757480. We also acknowledge support under the MIUR PRIN and FARE programmes (GW- NEXT, CUP: B84I20000100001).
\end{acknowledgments}

\bibliographystyle{utphys}
\bibliography{Ref}

\end{document}